\documentstyle[aps,prl,preprint,epsfig]{revtex}

\def\ket#1{{|#1\rangle}}
\def\bracket#1#2{{\langle #1 | #2 \rangle}}

\def\psum{{p_{\rm sum}}}
\def\tauabc{{\tau_{ABC}}}

\begin{document}

\draft

\tighten

\title{Parametrization and distillability of three-qubit entanglement}

\author{Todd A. Brun\thanks{Current address:  Institute for Advanced Study,
Einstein Drive, Princeton, NJ 08540.  Email:  tbrun@ias.edu} and
Oliver Cohen\thanks{Email:  ocohen@andrew.cmu.edu} \\
Physics Department, Carnegie Mellon University, \\
Pittsburgh,  PA  15213}

\maketitle

\begin{abstract}
There is an ongoing effort to quantify entanglement of quantum pure states
for systems with more than two subsystems.
We consider three approaches to this problem
for three-qubit states:  choosing a basis 
which puts the state into a standard form, enumerating ``local invariants,''
and using operational quantities such as the number
of maximally entangled states which can be distilled.  In this paper
we evaluate a particular standard form, the {\it Schmidt form}, which is
a generalization of the Schmidt decomposition for bipartite pure states.
We show how the coefficients in this case can be parametrized in terms
of five physically meaningful local invariants; we use this form
to prove the efficacy of a particular distillation technique for
GHZ triplets; and we relate the yield of GHZs to classes of states with
unusual entanglement properties, showing that these states represent
extremes of distillability as functions of two local invariants.
\end{abstract}

\pacs{03.67.-a 03.65.-w 03.67.Dd}

\section{Introduction}

The importance of {\it quantum entanglement}, 
both as a resource for quantum information processing and
as a ubiquitous feature of quantum systems, has become 
increasingly apparent over the last few years
\cite{Bennett93,Deutsch96,Bennett96a}.  Recent developments in 
quantum information theory, in particular, have stimulated interest in 
the quantification and manipulation of entanglement.

For bipartite pure states an essentially
complete theory of entanglement now exists
\cite{Bennett96a,Nielsen99,Approximate99},
though the situation for mixed states is less definite \cite{Horodecki96}.
All descriptions of bipartite pure state entanglement start with the
{\it Schmidt decomposition}.  It is possible to find
orthonormal bases $\{\ket{i}_A\}$ and $\{\ket{i}_B\}$ for systems A and B
such that we can write the joint state of the system in the form
\begin{equation}
\ket{\Psi_{AB}} = \sum_i \sqrt{p_i} \ket{i}_A \otimes \ket{i}_B, \ \ 
p_i > 0,\ \ \sum_i p_i = 1.
\label{schmidt_decomp}
\end{equation}
These {\it Schmidt coefficients} $\{p_i\}$ are uniquely defined by the state
$\ket{\Psi_{AB}}$, and are equal to the eigenvalues of the reduced
density matrix $\rho_A$ (or equivalently, of $\rho_B$); the bases
$\{\ket{i}_A\}$ and $\{\ket{i}_B\}$ are eigenbases of $\rho_A$ and
$\rho_B$, respectively, so the local density matrices are diagonal
in this choice of bases.  This choice of bases also minimizes the
number of terms needed to represent $\ket{\Psi_{AB}}$.

For tripartite or multipartite states, there is no equivalent to
the Schmidt decomposition (\ref{schmidt_decomp}) \cite{Thapliyal99}).
Three main approaches to parametrizing tripartite or multipartite
entanglement have been followed so far.
First, one may choose the local bases
to put the joint state into a {\it standard form}.
Often these standard forms are intended to
generalize some aspect of the Schmidt decomposition in the bipartite case
\cite{Linden98,Schlienz,Cohen,CohenBrun00,Sudbery00,Acin00,Higuchi00,CHS00}.
Second, one may try to identify a complete
set of {\it locally invariant quantities}, functions of the state
which are invariant under local unitary transformations
\cite{Linden98,Sudbery00,Linden99a,Kempe99,Coffman99,CarteretSudbery00},
and which uniquely characterize equivalent states.
The coefficients of a standard form are obviously such quantities, but
they may not have readily meaningful physical
interpretations.  Third, one may identify {\it operational quantities},
such as the number of Greenberger-Horne-Zeilinger (GHZ) triplets
or EPR pairs that can be distilled from the state by some procedure
\cite{CohenBrun00,Acin00b,Bennett99}.

In section II we consider a number of proposals for standard forms of
three-qubit pure states, concentrating especially on those which
generalize some aspect of the bipartite Schmidt decomposition:  the
{\it minimal} form \cite{Acin00,Higuchi00,CHS00},
the {\it two-term} form \cite{Acin00,Dur00,Acin00b},
and the {\it Schmidt} form.  This form was given briefly in \cite{CohenBrun00}
and independently in \cite{Sudbery00}.  In section III we examine it
in greater detail.  We give an explicit parametrization of the coefficients
in terms of five locally invariant quantities, and discuss their
physical significance.

We make use of the Schmidt form to prove analytically the reliability 
of a proposed distillation technique for GHZ triplets from general
three-qubit pure states \cite{CohenBrun00};
we present this proof in section IV.

In \cite{Linden98} Linden and Popescu proposed characterizing the
entanglement properties of three-qubit states by examining the
``orbits'' of the states under general local unitary transformation.
This was carried a step further by Carteret and Sudbery
\cite{CarteretSudbery00}, who proved that most states have a certain
generic behavior under such transformations, but identified
classes of `special' states which they speculated to have
unusual entanglement properties.  In section V we briefly review
these `special' states, then analytically evaluate the yield of GHZs
under the distillation protocol of \cite{CohenBrun00} and
section IV.  By examining the yield of these states as a function of
the invariant parameters from section III, and also of the locally invariant
``residual tangle'' $\tauabc$ of
Coffman, Kundu and Wootters \cite{Coffman99}, we verify that these
classes of states are indeed exceptional by this operational measure,
representing extremes of distillability or undistillability.  We briefly
compare the results using this protocol to the
recently discovered optimal distillation method of \cite{Acin00b},
and find that they are entirely consistent.
Our conclusions are summarized in section VI.

\section{Review of standard forms for three-qubit states}

Two qubits can always be represented in their Schmidt decomposition
(\ref{schmidt_decomp})
\begin{equation}
\ket\psi = \sqrt{p}\ket{00} + \sqrt{1-p}\ket{11},
\label{two_bit_schmidt}
\end{equation}
characterized by a single Schmidt coefficient $p$ (or equivalently
$1-p$).  Without loss of generality, we adopt the convention
that $p \ge 1/2$ and the corresponding eigenvector is $\ket0$.

Most attempts to define a standard form for a {\it three}-qubit state
attempt either to generalize some  property of the bipartite Schmidt
decomposition, or make use of the Schmidt decomposition between one
of the bits and the other two, or both.
For instance, we can make a Schmidt decomposition
between qubit A and qubits B and C, writing the three-qubit state in
the form
\begin{equation}
\ket\psi = \sqrt{p}\ket{0}_A\ket{\psi_0}_{BC}
  + \sqrt{1-p}\ket{1}_A\ket{\psi_1}_{BC}.
\label{partial_schmidt}
\end{equation}
Choosing the Schmidt basis for qubit A guarantees that the correlated
states of qubits B and C must be orthogonal:
$\bracket{\psi_0}{\psi_1} = 0.$

The sixteen real parameters to describe a generic pure state of three qubits
can be reduced to fifteen by normalization, and to five which are invariant
under the ten-dimensional group of local unitary transformations
\cite{Linden98,CarteretSudbery00}; unfortunately, no single
choice of five quantities has proven completely satisfactory.
One simple parametrization that has been proposed 
\cite{Linden98,Schlienz,Cohen} is the Linden-Popescu-Schlienz (LPS)
standard form.  One begins with
a state in form (\ref{partial_schmidt}).  One can then choose
one of the two correlated states, say $\ket{\psi_0}_{BC}$, and find
its corresponding Schmidt bases.  The resulting state for the three
qubits has the form
\begin{eqnarray}
\ket\psi &=& \sqrt{p} \ket{0}_A
  \left( a \ket{00}_{BC} + \sqrt{1-a^2} \ket{11}_{BC} \right) \nonumber\\
&& + \sqrt{1-p} \ket{1}_A \left(
  \gamma( \sqrt{1-a^2}\ket{00}_{BC} - a\ket{11}_{BC} )
  + f \ket{01}_{BC} + g \ket{10}_{BC} \right),
\end{eqnarray}
where $p$, $a$ and $f$ are real positive numbers, $g$ is complex, and
$\gamma = (1-f^2-|g|^2)^{1/2}$.  Together these give five independent
real parameters.

The vectors $\ket{\psi_0}_{23}$ and $\ket{\psi_1}_{23}$
span a two-dimensional subspace of the Hilbert space for qubits 2 and 3.
It's possible to make an interesting variation on the LPS idea
using a result of Niu and Griffiths, who showed
\cite{Niu99} that any such two-dimensional subspace can
be given basis vectors of the form
\begin{eqnarray}
\ket{\chi_0} &=& \sqrt{q}\ket{00}_{23} + \sqrt{1-q}\ket{11}_{23}, \nonumber\\
\ket{\chi_1} &=& \sqrt{r}\ket{01}_{23} + \sqrt{1-r}\ket{10}_{23}, 
\end{eqnarray}
for some choice of a product basis for the 4-D Hilbert space of the two bits,
where $q$ and $r$ are real numbers between 0 and 1.  Using this basis
leads to a unique standard form
\begin{eqnarray}
\ket\psi &=& \sqrt{p} \ket0
  \left( a \sqrt{q} \ket{00} + a \sqrt{1-q} \ket{11}
  + b \sqrt{r} \ket{01} + b \sqrt{1-r} \ket{10} \right) \\
&& + \sqrt{1-p} \ket1 \left( - b^* \sqrt{q} \ket{00} - b^* \sqrt{1-q} \ket{11}
  + a \sqrt{r} \ket{01} + a \sqrt{1-r} \ket{10} \right) , \nonumber
\end{eqnarray}
where $a$ is real and $a^2+|b|^2 = 1$.  This then gives
five independent real parameters: $p$, $q$, $r$, $a$, and the
phase of $b$.  This form treats
the $\ket0$ and $\ket1$ terms more symmetrically than LPS; however,
there is still a lack of symmetry under interchange of the bits.

More interesting from a fundamental point of view are attempts to
generalize some aspect of the Schmidt decomposition.  Three such
properties suggest themselves.  First, the Schmidt decomposition is
the choice of orthonormal bases for the local Hilbert spaces which
minimizes the number of terms needed to represent the state.
Second, any two qubit state can
be written as the sum of only two product vectors.  (For $N$-dimensional
systems, $N$ product vectors are needed.)  Third, the Schmidt
decomposition diagonalizes the reduced density matrices of the local
subsystems.  No single representation for tripartite systems has
all three properties, but they can be generalized individually.

Ac\'\i n et al. \cite{Acin00} have shown that all three-qubit states can be
written in the form
\begin{equation}
\ket\psi = \lambda_0 \ket{000} + \lambda_1 {\rm e}^{i\phi} \ket{100}
 + \lambda_2 \ket{101} + \lambda_3 \ket{110} + \lambda_4 \ket{111}
\label{minimal1}
\end{equation}
by a suitable choice of basis, where the $\lambda_i$ are all real and
positive and $\phi$ is a phase between $0$ and $\pi$.  With only
five terms, this is a minimal description, and in that sense a
generalization of the bipartite Schmidt decomposition.
A similar form has been described by Higuchi and Sudbery \cite{Higuchi00},
\begin{equation}
\ket\psi = \lambda_0 {\rm e}^{i\phi} \ket{000} + \lambda_1 \ket{100}
 + \lambda_2 \ket{010} + \lambda_3 \ket{001} + \lambda_4 \ket{111}
\label{minimal2}
\end{equation}
which has the added benefit of being symmetric under interchange of
the qubits.  Carteret, Higuchi and Sudbery \cite{CHS00} have shown
how to generalize this construction to give a unique minimal
representation for systems of any dimension.  These minimal forms
have practical benefits:  with a small number of terms, they can
simplify the calculation of locally invariant quantities \cite{Acin00c}.
However, the $\lambda_i$ and $\phi$ themselves have no obvious physical
interpretation.

This minimal property can be generalized in another way, by relaxing
the requirement that the product vectors be orthogonal.  Ac\'\i n et al.
and D\"ur, Vidal and Cirac have also shown \cite{Acin00,Dur00} that almost
all three qubit states can be written in the form
\begin{equation}
\ket{\Psi_{ABC}} = \mu_1 \ket{a_1b_1c_1}
  + \mu_2 {\rm e}^{i\phi} \ket{a_2b_2c_2},
\label{two_term}
\end{equation}
where the vectors are normalized but not orthogonal.
There are six real parameters,
$\mu_1$, $\mu_2$, $\bracket{a_1}{a_2}$, $\bracket{b_1}{b_2}$,
$\bracket{c_1}{c_2}$ and $\phi$; imposing normalization reduces
this to five.

Interestingly, not all three qubit states can be written in the
form (\ref{two_term}); a small subclass of states require a minimum
of three product terms \cite{Acin00}.  D\"ur, Vidal and Cirac made
use of this result to prove that there are two classes of three-qubit
pure states which cannot be interconverted with nonzero probability
\cite{Dur00}.  The class that requires three terms is a three-parameter
family, and is characterized by vanishing residual tangle
$\tauabc=0$ \cite{Coffman99} (see section V).
Ac\'\i n, D\"ur and Vidal also used
this form to demonstrate a method of converting
a single copy of a three qubit state into a GHZ triplet with maximum
probability \cite{Acin00b}.

The third generalization is to find bases for all three qubits
which diagonalize their reduced density matrices.  That is, one can
simultaneously put each bit in its Schmidt decomposition with respect
to the other two.  This form was proposed in \cite{CohenBrun00} and
independently in \cite{Sudbery00}.  The state has the form
\begin{eqnarray}
\ket\psi &=& a\ket{000} + b\ket{001} + c\ket{010} + d\ket{011} \nonumber\\
&& + e\ket{100} + f\ket{101} + g\ket{110} + h\ket{111},
\label{three_qubit}
\end{eqnarray}
which looks just like a generic three-qubit state with 16 parameters.
However, using each of the three qubits in turn
we can write $\ket\psi$ in a form similar to (\ref{partial_schmidt}),
with orthogonality conditions which
impose restrictions on the possible values of the coefficients
in (\ref{three_qubit}).  We can use these relationships to reduce
these coefficients to five independent parameters, as we show in
the next section.

\section{Parametrizing the Schmidt form}

By redefining the relative phases of the basis vectors
\begin{equation}
\ket0_j,\ket1_j \rightarrow \exp(i\phi_j)\ket0,\exp(i\theta_j)\ket1,
\end{equation}
we can choose to make four of the coefficients real.  A convenient choice
is to make $a,d,f,g$ real, while $b,c,e,h$ remain complex.  The state
must also be normalized, which imposes the condition
\begin{equation}
a^2 + |b|^2 + |c|^2 + d^2 + |e^2| + f^2 + g^2 + |h|^2 = 1.
\label{normalization}
\end{equation}
This leaves 11 undetermined parameters.

We can now express the larger eigenvalues $p_{A,B,C}$ of the reduced 
density matrices $\rho_{A,B,C}$ in terms of the coefficients:
\begin{equation}
p_A = a^2 + |b|^2 + |c|^2 + d^2, \ \ {\rm etc.,}
\label{p_equation}
\end{equation}
(the smaller eigenvalues obviously being $1-p_{A,B,C}$).
Finally, the states $\ket{\psi_{0,1}}_{kl}$
correlated with basis vectors $\ket0_j$ and $\ket1_j$
must be orthogonal to each other.  This gives three more equations:
\begin{equation}
a e^* + b f + c g + d h^* = 0, \ \ {\rm etc.}
\label{orthogonality}
\end{equation}
Because these equations are complex, they are equivalent to six real
equations.

Combining these restrictions, we now have fourteen equations in sixteen
unknowns.  Thus, in addition to the eigenvalues $p_{A,B,C}$ we would
expect there to be two more free parameters.  Can we identify reasonable
candidates for these parameters?
It turns out that natural choices are the two probabilities
$a^2$ and $|h|^2$.  These parameters are symmetric under interchanges of
the three qubits, and have a fairly simple physical interpretation:
they are the probabilities of all three qubits giving the same result
(0 or 1, respectively) when measured in their Schmidt bases.  Moreover,
the coefficients of the other state vectors can all be calculated in terms
of the five probabilities $a^2,|h|^2$, and $p_{A,B,C}$, up to a sign.

Define $\psum=p_A+p_B+p_C$.
The expressions for the norms of the coefficients are then simple:
\begin{eqnarray}
|b|^2,|c|^2,|e|^2 &=& { { (2p_{C,B,A} - 1)|h|^2
  - (\psum - p_{C,B,A} - 1)(2a^2-\psum+1) }
  \over{ 2\psum-3 } } \nonumber\\
d^2,f^2,g^2 &=& { { (2p_{A,B,C} - 1)a^2
  - (\psum - p_{A,B,C} - 1)(2|h|^2+\psum-2) }
  \over{ 2\psum-3 } } \;.
\label{norm_equations}
\end{eqnarray}

The phases of $b,c,e$ are more complicated.  If we define the variables
$\phi_{b,c,e}$ by $b=|b|\exp(i\phi_b)$, $c=|c|\exp(i\phi_c)$, and
$e=|e|\exp(i\phi_e)$, the constraint equations
(\ref{orthogonality}) imply after a bit of algebra that
\begin{eqnarray}
\cos(\phi_b), \sin(\phi_b)
  &=& (Q_{1,2}/|b|)(\mp 2adf + g(a^2+d^2+f^2-g^2) ), \nonumber\\
\cos(\phi_c), \sin(\phi_c)
  &=& (Q_{1,2}/|c|)(\mp 2adg + f(a^2+d^2-f^2+g^2) ), \nonumber\\
\cos(\phi_e), \sin(\phi_e)
  &=& (Q_{1,2}/|e|)(\mp 2afg + d(a^2-d^2+f^2+g^2) ), \nonumber\\
\cos(\phi_h), \sin(\phi_h)
  &=& (Q_{1,2}/|h|)(\mp 2dfg + a(-a^2+d^2+f^2+g^2) ),
\label{phase_equations}
\end{eqnarray}
where $Q_1$ and $Q_2$ are two constants.  We can solve for the values
of $Q_1$ and $Q_2$ by using the identity $\sin^2(\phi)+\cos^2(\phi)=1$
and substituting (\ref{norm_equations}) for $|b|,\ldots,g$.

In the Schmidt form for three-qubit pure states, each of the
five parameters has a reasonably straightforward physical 
interpretation.  The three parameters $p_A,p_B,p_C$ are the larger
(i.e., $p\ge1/2$) eigenvalues of the reduced density operators for each
of the three qubits, and correspond to the probabilities of
obtaining the more likely of the two possible outcomes (which by convention
we label $\ket0$) when we measure each of the qubits in its
Schmidt basis.  These parameters are closely related to the
minimum absolutely selective information \cite{CohenBrun00} for each 
qubit, which is given by the entropy function
\begin{equation}
\min S_i = -(p_i\log_2 p_i + (1-p_i)\log_2 (1-p_i)).
\end{equation} 
This quantity is the minimum amount of {\it fundamentally unpredictable}
classical information generated by carrying out a measurement on 
qubit $i$, given a free choice of measurement basis \cite{CohenBrun00}.
By using the Schmidt form to choose measurement bases we can simultaneously 
minimize the absolutely selective information for all three qubits.

The parameters $p_A,p_B,p_C$ range from $1/2$ to $1$ (since they
are defined to be the {\it larger} eigenvalues of their corresponding
local density matrices).  Similarly, $a^2$ ranges from $0$ to $1$,
and $|h|^2$ from $0$ to $1/2$.
However, this does not mean that
these parameters can take arbitrary values within these ranges.  Some
choices of parameter values correspond to no physical state, and
give nonsensical values for (\ref{norm_equations}) and
(\ref{phase_equations}).

In particular, the local probabilities must obey the triangle inequalities
\begin{eqnarray}
p_A(1-p_A) + p_B(1-p_B) &\ge& p_C(1-p_C), \nonumber\\
p_B(1-p_B) + p_C(1-p_C) &\ge& p_A(1-p_A), \nonumber\\
p_C(1-p_C) + p_A(1-p_A) &\ge& p_B(1-p_B);
\label{triangle}
\end{eqnarray}
these imply, for instance, that if $p_A=1$ then $p_B=p_C$.  The
restrictions on $a^2$ and $|h|^2$ are more complicated, but they too
display an interdependency in their range of values.  In particular,
as $\psum\rightarrow3$ we must have $a^2\rightarrow1$ and
$|h|^2\rightarrow0$.

\section{Proof of distillability}

The Schmidt form can provide analytical insight when 
addressing specific problems. For example, the efficacy of a recently 
proposed tripartite distillation protocol \cite{CohenBrun00} can be
demonstrated with its help.

Consider a state of three qubits in an arbitrary product basis,
which can be written in the form (\ref{three_qubit}).  We can
straightforwardly calculate the quantity $p_A(1-p_A)$
\begin{eqnarray}
p_A(1-p_A) &=& |af-be|^2 + |ag-ce|^2 + |ah-de|^2 \nonumber\\
&& + |bg-cf|^2 + |bh-df|^2 + |ch-dg|^2,
\label{p-p2}
\end{eqnarray}
This expression is a polynomial in the coefficients and their
complex conjugates, and is correct in any basis.  If the state is in
the Schmidt form, this simplifies to
\begin{equation}
p_A(1-p_A) = (a^2 + |b|^2 + |c|^2 + d^2)(|e|^2 + f^2 + g^2 + |h|^2).
\label{Schmidt_p-p2}
\end{equation}

Let us assume that we have written the state in Schmidt form,
such that the states $\{\ket0,\ket1\}$ for each qubit $j$ are eigenstates
of the local density matrix with eigenvalues $p_j$ and $1-p_j$, respectively.
Suppose we now perform a weak measurement on each of the three qubits.
First, allow each qubit to interact with a separate ancilla bit initially
in state $\ket0$, such that
\begin{eqnarray}
\ket0 \otimes \ket0_{\rm anc} &\rightarrow&
  \sqrt{1-\epsilon} \ket0 \otimes \ket0_{\rm anc}
  + \sqrt{\epsilon} \ket0 \otimes \ket1_{\rm anc}, \nonumber\\
\ket1 \otimes \ket0_{\rm anc} &\rightarrow&
  \ket1 \otimes \ket0_{\rm anc},
\end{eqnarray}
where $\epsilon \ll 1$.  Then measure the three ancilla bits.  With a
probability of $\epsilon(\psum)$ one will find one or
more of the ancilla bits in state $\ket1_{\rm anc}$, in which case
the procedure has failed.  Otherwise, this step has succeeded and the
three qubits are now in a new state with slightly different coefficients
$a',b',\ldots,h'$.  The changes in the coefficients are
\begin{eqnarray}
\Delta a &=& - (\epsilon/2) (3 - \psum) a, \nonumber\\
\Delta (b,c,e)
  &=& - (\epsilon/2) (2 - \psum) (b,c,e), \nonumber\\
\Delta (d,f,g) &=& - (\epsilon/2) (1 - \psum) (d,f,g), \nonumber\\
\Delta h &=&  (\epsilon/2)  \psum h.
\label{Delta_coeff}
\end{eqnarray}
This very simple form results because the state is in Schmidt form.
After this procedure the bases for the three bits will generally no longer
be the correct Schmidt basis (though it will be close to it), so the
expression (\ref{Schmidt_p-p2}) cannot be used; but (\ref{p-p2})
is always correct.  Thus we get a change in $p_A(1-p_A)$
\begin{eqnarray}
\Delta[p_A(1-p_A)] &=& 
 - \epsilon (4 - 2(p_A+p_B+p_C)) (|af-be|^2 + |ag-ce|^2) \\
&& - \epsilon (3 - 2(p_A+p_B+p_C)) (|ah-de|^2 + |bg-cf|^2) \nonumber\\
&& - \epsilon (2 - 2(p_A+p_B+p_C)) (|bh-df|^2 + |ch-dg|^2) \nonumber\\
&=& - \epsilon (3 - 2(p_A+p_B+p_C)) p_A(1-p_A) \nonumber\\
&& - (\epsilon/2)(|af-be|^2 + |ag-ce|^2 - |bh-df|^2 - |ch-dg|^2 ) \nonumber
\end{eqnarray}
By making use of equations (\ref{p_equation}) and (\ref{orthogonality}),
this expression simplifies to
\begin{eqnarray}
\Delta[p_A(1-p_A)] &=&
 \epsilon \biggl[ (2(p_A+p_B+p_C) - 3) p_A(1-p_A) \nonumber\\
&& + p_A (a^2 - |e|^2 + |h|^2 - d^2 ) + d^2 - a^2 \biggr],
\end{eqnarray}
which using (\ref{norm_equations}) further simplifies to
\begin{eqnarray}
\Delta[p_A(1-p_A)] &=&
 {{\epsilon (2p_A-1)}\over{2p_A+2p_B+2p_C-3}} \biggl[ 2(a^2+|h^2|)(p_B+p_C-1)
  \nonumber\\
&& - (2p_A-1)(p_A+p_B+p_C-1)(p_A+p_B+p_C-2)  \biggr].
\label{Deltap-p2}
\end{eqnarray}
The prefactor to (\ref{Deltap-p2}) is strictly positive, as is the
first term inside the brackets.  The second term is positive if
$p_A+p_B+p_C<2$; any state that satisfies this criterion will evolve
towards the GHZ state and have a nonzero yield.

For $p_A+p_B+p_C \ge 2$, the sign of (\ref{Deltap-p2}) depends on the
relative sizes of the first and second terms inside the brackets.
The last two equations of (\ref{norm_equations}) show that
for $p_A+p_B+p_C \ge 2$, the fact that $f^2+g^2>0$ implies
\begin{equation}
2 a^2 (p_B+p_C-1) \ge (2p_A+p_B+p_C-2)(p_A+p_B+p_C-2),
\end{equation}
which yields the inequalities
\begin{eqnarray}
&& 2a^2(p_B+p_C-1) - (2p_A-1)(p_A+p_B+p_C-1)(p_A+p_B+p_C-2) \nonumber\\
&\ge& (2p_A+p_B+p_C-2)(p_A+p_B+p_C-2) \nonumber\\
&&  - (2p_A-1)(p_A+p_B+p_C-1)(p_A+p_B+p_C-2) \nonumber\\
&=&  (1-p_A)(p_A+p_B+p_C-2)(2p_A+2p_B+2p_C-3) \ge 0.
\end{eqnarray}
This straightforwardly implies
\begin{equation}
\Delta[p_A(1-p_A)] \ge
  \epsilon (2p_A-1)(1-p_A)(p_A+p_B+p_C-2) \ge 0.
\end{equation}
Because of the symmetry of the protocol, $p_B(1-p_B)$ and $p_C(1-p_C)$
must also increase.  So one step of this protocol must move the state
towards the GHZ with nonvanishing probability, and will (in general)
produce a nonzero yield of GHZ triplets.

There are three circumstances in which this result can fail.  First,
no product state can ever be distilled to a GHZ by this method.  At least
one of $p_A,p_B,p_C$ must equal 1 in this case, which causes the rate
(\ref{Deltap-p2}) corresponding to it to vanish.  This is not immediately
obvious from the form of (\ref{Deltap-p2}), but it is easily checked
using (\ref{norm_equations}) and (\ref{triangle})---if $p_A=1$,
then $p_B=p_C=a^2$, and
(\ref{Deltap-p2}) is equal to zero.

Second, there are states with
$p_A+p_B+p_C=2$ for which $a^2=|h|^2=0$, again making (\ref{Deltap-p2})
vanish.  These are a subset of the {\it triple states} discussed
in section V below, which are equivalent to states 
of the form (\ref{triple}); these states have vanishing residual tangle.
Finally, it is possible for a state
with $p_A+p_B+p_C>2$ to evolve to one of these triple states.
All such states will also have vanishing residual
tangle \cite{Dur00}, and conversely all states with vanishing
residual tangle will evolve under this distillation protocol to a
triple state with $p_A+p_B+p_C=2$, and hence have zero yield of GHZs.
This can be clearly seen in Fig.~1.

\section{Entanglement and distillability}

Linden and Popescu \cite{Linden98} proposed characterizing
three-qubit states by the dimensions of their orbits under the
action of the local unitary group.  Generically, tripartite pure
states of qubits have ten-dimensional orbits, equal to the dimension
of the local unitary group.
The very interesting results of Carteret and Sudbery
\cite{CarteretSudbery00} give a complete classification of all
states for three qubits which behave nongenerically under
local unitary transformations; these `special' states have
stabilizers of nonzero dimension, and hence orbits of dimension
$<10$ (see \cite{CarteretSudbery00}).
This behavior suggests that these `special' classes have unusual
entanglement properties, which might be evident in other
measures of entanglement.

We have numerically simulated the distillation of generic states
by the algorithm described above in section IV, in order to determine
the yield of GHZ triplets as a function of various parameters, especially
the parameters used to describe the Schmidt form.  We
have also calculated analytical expressions for the yield of states
in the exceptional classes enumerated by Carteret and Sudbery.  We
find that these states are indeed exceptional by this operational
criterion, as we describe below.

Most important in calculating the yield of GHZs is the sum of the
local eigenvalues $\psum \equiv p_A+p_B+p_C$.  This quantity determines
the probability of failure in one step of the infinitesimal distillation
procedure of section IV, with the probability of failure being
$\epsilon\psum$.  If it takes $N$ steps to become sufficiently close
to a GHZ triplet, the expected yield is
\begin{equation}
Y = \prod_{n=1}^N (1 - \epsilon\psum(n))
  \approx \exp\left\{ - \sum_n \epsilon\psum(n) \right\},
\label{yield_integral}
\end{equation}
where $\psum(n)$ is the value of $\psum$ at the $n$th step.  In the limit
of infinitesimal steps the sum inside the exponent becomes an integral.
Calculating $\psum(n)$ analytically is no simple matter for a general
state; the equations (\ref{Delta_coeff}) for the change in the coefficients
become differential equations in the limit, but must be supplemented
by an additional change of basis between steps, since in general
the bases will no longer be the Schmidt basis for the new state.
While this is simple to do numerically, analytically it is challenging.

Fortunately, the classes of exceptional states are generally expressible
in simple forms which make it possible to integrate the equations
(\ref{Delta_coeff}) in closed form, and derive simple expressions for
the yield of GHZs.
Interestingly, the steps of the GHZ distillation technique commute with
local unitary transformations.  Because of this, the distillation procedure
preserves the stabilizer of the initial state, and hence must take
`special' states to other `special' states of the same type.  This
gives another way of understanding why certain special states are not
distillable to GHZs.

In addition to $a^2$, $|h|^2$, and $\psum$, we looked at the
dependence of the GHZ yield on one other locally invariant quantity.
This is the {\it residual tangle} of
Coffman et al. \cite{Coffman99}, which can be written
\begin{equation}
\tauabc = 2(\lambda_1^{AB}\lambda_2^{AB}+\lambda_1^{AC}\lambda_2^{AC}),
\end{equation}
where 
$\lambda_1^{ij}$ and $\lambda_2^{ij}$ are the (positive) eigenvalues of the 
matrix $\sqrt{\rho_{ij}\tilde\rho_{ij}}$. Here $\rho_{ij}$  is the 
density operator for the two-party $ij$ system, and $\tilde\rho_{ij}$ 
is the ``spin-flipped'' density operator: $\tilde\rho_{ij}=(\sigma_y 
\otimes \sigma_y)\rho_{ij}^*(\sigma_y \otimes \sigma_y)$.
It has been suggested \cite{Coffman99}
that the residual tangle is a measure of
the irreducible three-way (``GHZ-type'') entanglement of a tripartite state,
beyond any two-party (``EPR-type'') entanglement that may be contained
in such a state.  As such, it is of particular interest in discussing
distillability below.  Also, its square $\tauabc^2$ is a 
polynomial quantity, which makes it analytically tractable.

{\it Triple States.}  For this set of states the residual tangle 
vanishes \cite{Coffman99}.  We previously described
states in this set as ``triple'' states \cite{CohenBrun00},
because they are equivalent under local unitary transformations
to states with just three components:
\begin{equation} 
\ket{\psi_{\rm tr}} = b \ket{001} + c \ket{010} + e \ket{100}.
\label{triple}
\end{equation} 
Carteret and Sudbery \cite{CarteretSudbery00}
refer to these as ``beechnut'' states; they all have
$\psum\ge2$.
For triple states with $\psum>2$ each step of the infinitesimal
distillation protocol reduces $\psum$, but leaves the state a triple
state.  If $\psum=2$, the actions on the three
qubits cancel out, leaving the state unchanged.
States of this type have vanishing primary yield for the tripartite
distillation protocols described in section IV and in \cite{CohenBrun00};
indeed, D\"ur, Vidal and Cirac have shown that {\it no} procedure can
transform one copy of a state with zero residual tangle into
a GHZ with nonzero probability \cite{Dur00}.

Because the distillation procedures of section IV and \cite{CohenBrun00}
preserve the classes of `special' states, it is easy to see why they
cannot produce GHZs from triple states; because 
all triple states have $\psum\ge2$,
they cannot include the GHZ state ($\psum=3/2$) as a limit.
The set of product states (or ``bystander states'' in the terminology of
Carteret and Sudbery) is similarly undistillable.  The result of
D\"ur, Vidal and Cirac, however, goes beyond this, since it assumes
nothing about the symmetry of the procedure.

The symmetric version of state (\ref{triple}) (with $b=c=e=1/\sqrt{3}$)
is termed by D\"ur, Vidal and Cirac the ``W'' state, and seems to fill
a role for the zero residual tangle states similar to the role filled
by the GHZ for all other states:  it is, in some sense, maximally
entangled.  We will say a bit more about this below.
All other `special' classes include the GHZ as a limit,
and therefore are distillable.

{\it Generalized GHZ states.}  These states can be written in Schmidt form
\begin{equation}
\ket\psi = a\ket{000} + h\ket{111}.
\label{generalized_ghz}
\end{equation}
They have $p_A = p_B = p_C = a^2$, residual tangle $\tauabc = 4a^2 h^2$.
A single step of the infinitesimal distillation procedure gives a
new generalized GHZ with coefficients $a' = a + \Delta a$,
$h' = h + \Delta h$:
\begin{equation}
\Delta a = - (\epsilon/2)(3-\psum) a,\ \ 
\Delta h = + (\epsilon/2) \psum h,
\end{equation}
so (\ref{yield_integral}) can readily be evaluated to give
the yield of GHZs
\begin{equation}
Y = 1-\sqrt{1-\tauabc} = (2/3)(3-\psum).
\end{equation}
These states are the most distillable three qubit states as a function
of both $\tauabc$ and $\psum$; we can see this in Figures 1 and 2 below.

{\it Slice states.} In Schmidt form these are
\begin{equation}
\ket\psi = a \ket{000} + d \ket{011} - e\ket{100} + h \ket{111},\ \ 
ae=dh,
\label{slice}
\end{equation}
plus similar states derived by permuting the order of the bits.
These states have $p_B=p_C=a^2+e^2$, $p_A=a^2+d^2$,
$\tauabc = 4(ah+de)^2 = 4a^2 (h+e^2/h)^2$.
Imposing normalization and the orthogonality condition on (\ref{slice})
we see that this is a two-parameter family of states.  For these
two parameters we may choose $a^2$ and $h^2$, or equivalently
$p_A$ and $p_B$.
One step of the infinitesimal distillation protocol applied to state
(\ref{slice}) leaves qubits B and C in their Schmidt bases, but
not qubit A; a change of basis must be applied to A to put the new state
in Schmidt form.  This new state is still a slice state, and has
new parameters $p_A' = p_A + \Delta p_A$, $p_B' = p_B + \Delta p_B$,
\begin{eqnarray}
\Delta p_A &=& 2 \epsilon (p_A + 2 p_B - 1) p_A,  \nonumber\\
\Delta p_B &=& 2 \epsilon (p_A + 2 p_B - 2) p_B
  - \epsilon p_A(p_A+p_B-1)/(2p_A-1).
\end{eqnarray}
The yield is difficult to evaluate analytically, but numerical
evidence shows that generic slice states are not extremes of
distillability.  With each step of the distillation protocol, the
parameters $p_B=p_C$ approach $1/2$, but $p_A$ actually moves away.
However, when $p_B=p_C=1/2$, this subclass of slice states {\it does}
have extremal behavior.  Carteret and Sudbery term this subclass
the {\it maximal} slice states.

{\it Maximal slice} or {\it Slice-ridge states} are of form
(\ref{slice}) with $a^2+e^2 = 1/2 = p_B = p_C$.
This subclass is parametrized by a single number, which can be taken
to be $p_A$.  Because only $p_A$ is larger
than $1/2$, there is no need to perform the GHZ distillation procedure
on qubits B and C; performing it on A alone preserves the form of
the state, with
\begin{equation}
\Delta p_A = - (\epsilon/2)p_A(1-p_A),
\end{equation}
giving a yield of GHZs
\begin{equation}
Y = 1-\sqrt{1-\tauabc} = 2(2-\psum).
\end{equation}
The expression for the primary yield in terms of the residual tangle
is identical to that for the GHZ-type states, while 
in terms of $\psum$ it is not.  In terms of $\tauabc$ it is one
of the most distillable types of state (see Fig. 1).  In terms
of $\psum$ (Fig. 2) it appears to be one of the {\it least} distillable
types of states; this is because maximal slice states have the minimum 
$\tauabc$ of all states with a given $\psum$.

In addition to these `special' states, there are two classes of states
that deserve additional attention.  While these states have stabilizers
of zero dimension \cite{Carteret} like generic states, these classes
are also preserved by the above distillation protocols.
Like the `special' states, they extremize
distillability as a function of $\tauabc$ and $\psum$.

{\it Generalized triple} or {\it Tetrahedral states.}
These states can be written 
\begin{equation}
\ket\psi = b \ket{001} + c \ket{010} + e  \ket{100} + h \ket{111}.
\label{generalized_triple}
\end{equation}
We are mainly interested here in the symmetric state $b=c=e$; for this
case $p_A=p_B=p_C=2b^2$, $\psum\le2$.  This form is preserved by
the steps of the infinitesimal distillation protocol, which make the
coefficients evolve according to (\ref{Delta_coeff}); the yield is easily
integrated according to (\ref{yield_integral}) to give
$Y=2(2-\psum)=4(1-3b^2)$;
the residual tangle is $\tauabc=16 b^3\sqrt{1-3b^2} = \sqrt{(4-Y)^3 Y/27}$.
This yield is identical to that of the maximal slice states as a function
of $\psum$, but not as a function of $\tauabc$; from Figs. 1 and 2, we
see that they are states of minimal distillability in terms of both
$\psum$ and $\tauabc$.

{\it Zero residual tangle (ZRT) states.}  D\"ur, Vidal and Cirac have
shown that no states with $\tauabc=0$ can be converted to GHZ triplets
with nonzero probability, so $Y=0$.
They also showed that all such states can be written in the form
\begin{equation}
\ket\psi = a \ket{000} + b \ket{001} + c \ket{010} + e \ket{100}.
\label{zrt}
\end{equation}
This is in general not in the Schmidt form of section III.
These states include the triple states $a=0$ as a subclass (for
which (\ref{zrt}) {\it is} in Schmidt form).  The triple states form
a boundary of this set, and any ZRT state will evolve under the
distillation protocol to a triple state.  These states have
$\psum \ge 2$.

All these `special' states have symmetries which
account for both their enlarged stabilizers and their extremal
distillability.  One way of seeing this is to note that the various
standard forms given in section II, which for generic states all require
distinct bases, often coincide for these special states.  For instance,
the generalized GHZ states (\ref{generalized_ghz})
are simultaneously in Schmidt, two-term,
minimal, LPS {\it and} Griffiths-Niu form.  ZRT and triple states cannot
be written in two-term form, but can be written with three terms; the
triple states (\ref{triple}) are simultaneously in Schmidt, minimal,
three-term and Griffiths-Niu form.
Slice states written in the form (\ref{slice}) are simultaneously
in both Schmidt and LPS standard forms.

It is easiest to see how the the distillability of these states compares
to that of generic states by plotting their yields $Y$  as a function of
$\psum$ and $\tauabc$ along with the numerical results for a large 
sample of randomly generated states.  We have plotted these quantities
in Figs.~1 and 2, with the families of `special' states indicated.
We see that most of these states are indeed special as far as distillation
is concerned:  they form the boundaries of the plotted regions.
The quantity $\tauabc$ does seem to be closely related to distillability,
as conjectured, though this relationship is not exact; for
a given value of $\tauabc$ states with a range of $Y$ values exist,
but the range is not very wide.  This range is bounded at the top
by the generalized GHZ and maximal slice states, and at the bottom
by the symmetric generalized triple state.  All ZRT states
have $Y = \tauabc = 0$.  There is also a relationship between
$\psum$ and $Y$, though again for a given $\psum$ there is
a range of $Y$ values.  This range too is bounded above by the generalized
GHZs, and below by the ZRT states, generalized triples and maximal
slice states.  These upper and lower bounds are both linear; the upper
bound is exactly the same as that for Bernstein and Bennett's Procrustean
technique \cite{Bennett96a}, reflecting the fact that generalized GHZ states
can be distilled by exactly the same techniques which work in the
bipartite case.

A reasonable question is to what extent these yields are artifacts of
the particular distillation protocol we use.  After all, this technique
is only one possible way of producing GHZs, in general not the optimal
method even for a single copy of a three-qubit state.

Fortunately, we can actually answer this question.  Recently,
Ac\'\i n, Jan\'e, D\"ur and Vidal \cite{Acin00b} have discovered
the optimal algorithm for transforming a single copy of a three-qubit
state into a GHZ.  This involves performing a POVM on each of the three
bits, designed to project the states in the two-term representation
onto tri-orthogonal vectors.  Finding the correct POVM for an arbitrary
state involves maximizing a somewhat involved function, but is easily
done numerically.  We have done so for a large sample of random
states, as well as for the members of the `special' classes enumerated
by Carteret and Sudbery.

The optimal yield is higher, in general, than that of the infinitesimal
algorithm of section IV, though they are surprisingly close for most
states.  However, for the `special' states, the yields are identical.
In other words, the infinitesimal distillation technique gives the
optimal yield for these classes of states. Quite remarkably, if we
plot Figures 1 and 2 for the optimal GHZ distillation protocol, the
figures look completely unchanged.

Thus we can see that by both the optimal and the infinitesimal techniques,
these classes of special states extremize the yield of GHZs as a function
of both $\tauabc$ and $\psum$.  This strongly supports the conclusion that
these states do indeed have unusual entanglement properties, and are
worthy of further study.

\section{Conclusions}

We have examined tripartite entanglement from both an analytical and
an operational point of view.  In the bipartite case, which is well
understood and to which we have turned for clues, the analytical and
operational aspects of entanglement are closely related:
the entanglement properties of a single copy are given by the locally
invariant parameters, the Schmidt coefficients, which also determine their
operational characteristics.  We have looked for similar connections in the
three-qubit case.  Here at least five locally invariant parameters are
required, as opposed to just one in the two-qubit case.
We have examined several ways of choosing these five parameters, looking
in particular at generalizations of the bipartite Schmidt decomposition.
One representation in particular, the ``Schmidt form,'' has useful
properties which made it simple to prove the efficacy of the infinitesimal
GHZ distillation protocol of \cite{CohenBrun00}; it can also be
parametrized in terms of five physically meaningful quantities.

We have looked for connections between these parameters and yields in
distilling GHZ triplets, as well as connections with the residual
tangle of Coffman et al.  We have shown that the
`special' classes of states enumerated by the theorem of Carteret and Sudbery
extremize the distillation yield as functions of
the residual tangle $\tauabc$ and $\psum = p_A+p_B+p_C$.

Although a certain amount amount of progress towards understanding
tripartite entanglement has been made, at least for qubits,
many important questions remain unanswered.  For example, the number of
states in the asymptotic minimum reversible entanglement generating
set (MREGS) \cite{Bennett99,Linden99b,WuZhang00} for three-qubit states,
and for tripartite states in general, is still unknown.  No asymptotically
reversible  (or optimal but irreversible) distillation technique for GHZ
states is known.  The search for solutions to these and related problems
is ongoing.

\section*{Acknowledgments}

We would like to thank H.A. Carteret, W. D\"ur,  R.B. Griffiths,
A. Sudbery and G. Vidal for many useful conversations.
This work was supported by NSF Grant No. PHY-9900755.

\vfil\eject\vfil

Figure 1.  Here we plot the primary yield of GHZ triplets from the
infinitesimal distillation algorithm of section III vs. the square
of the residual tangle $\tauabc^2$
for various `special states' as well as a random sample of
generic states.  We see that all states lie between two curved boundaries;
the generalized GHZ and maximal slice states lie on the upper boundary, while
the generalized triple states lie on the lower boundary.  The triple
states all have both $\tauabc$ and the yield equal to zero.
Interestingly, the maximal slice states appear to be high-yield states when
plotted against $\tauabc$, but low-yield when plotted against
$p_{\rm sum} = p_A+p_B+p_C$; for a given value of $\tauabc$ these
states minimize $p_{\rm sum}$.
\vfil

Figure 2.  Here we plot the primary yield of GHZ triplets from the
infinitesimal distillation algorithm of section III vs. $p_{\rm sum} =
p_A+p_B+p_C$ for various `special states' as well as a random sample of
generic states.  We see that all states lie between two linear boundaries;
the generalized GHZ states lie on the upper boundary, while the maximal
slice and generalized Triple states lie on the lower boundary, and the
triple states are the zero-yield states between $p_{\rm sum} = 2$ and
$p_{\rm sum} = 3$.  The upper linear boundary corresponds to the yield
of Bernstein and Bennett's Procrustean method of EPR distillation
in the bipartite case.
\vfil

\begin{figure}[t]
\begin{center}
\epsfig{file=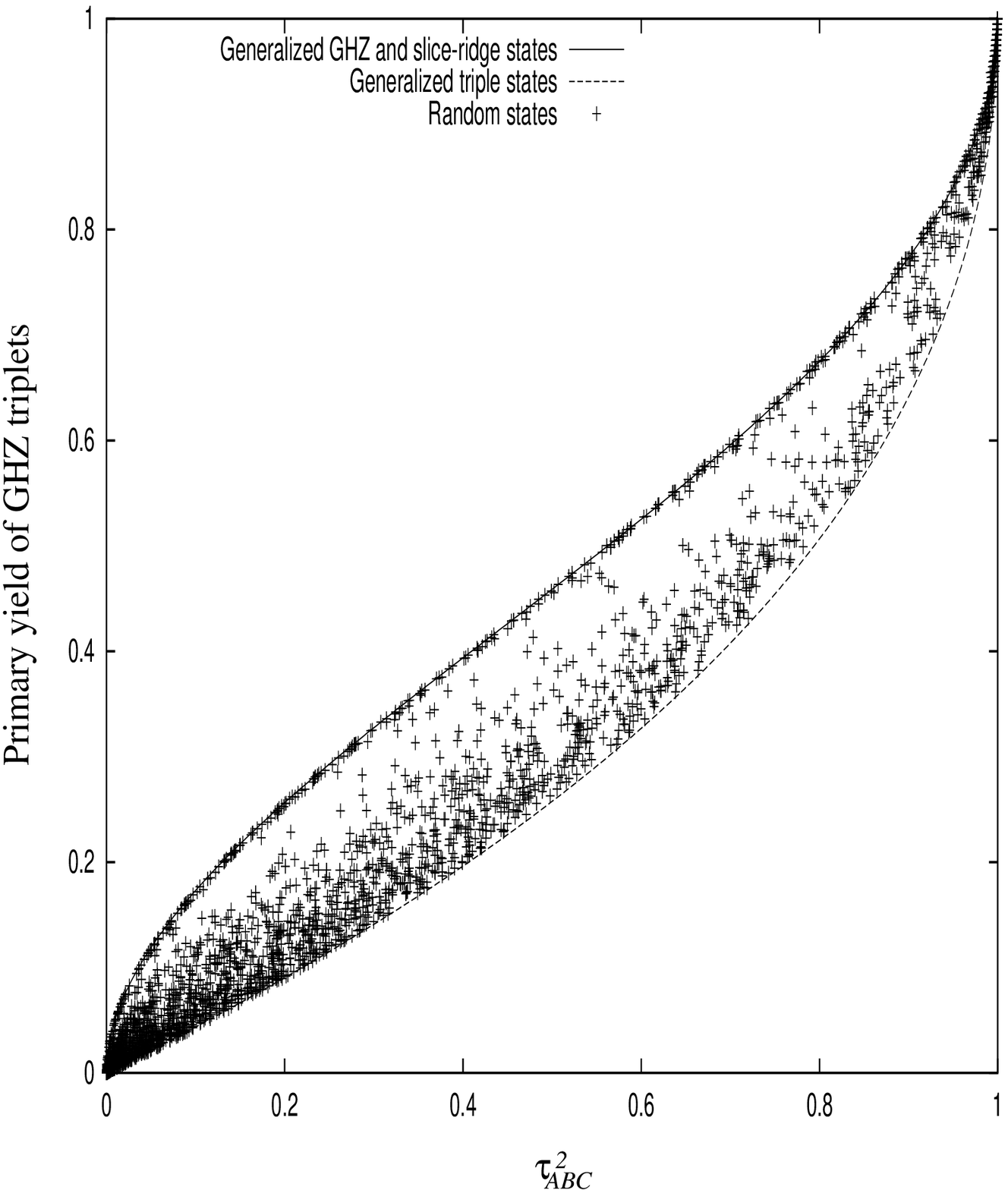}
\label{fig1}
\end{center}
\end{figure}
\centerline{Figure 1.}

\begin{figure}[t]
\begin{center}
\epsfig{file=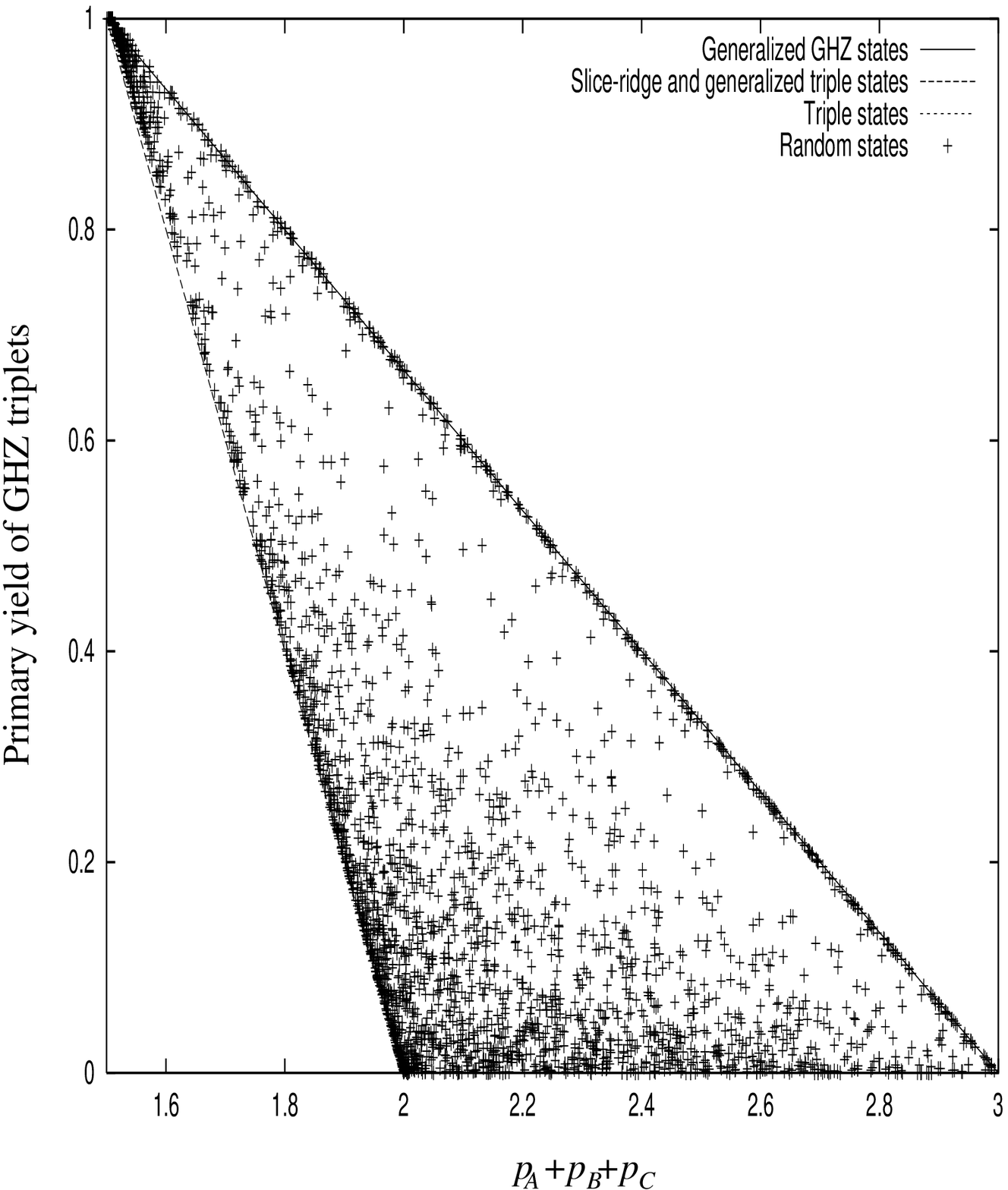}
\label{fig2}
\end{center}
\end{figure}
\centerline{Figure 2.}

\end{document}